\documentclass[twocolumn,aps,prd,unsortedaddress,%
superscriptaddress,a4paper,nofootinbib,balancelastpage,preprintnumbers,floatfix]{revtex4-1}
\usepackage{latexsym,amsbsy,amsmath}
\usepackage{graphicx}
\usepackage[utf8]{inputenc}
\mathchardef\mhyphen="2D
\begin{document}
\title{Very heavy flavored dibaryons}
\author{Jean-Marc~Richard}
\email{j-m.richard@ipnl.in2p3.fr}
\affiliation{Universit\'e de Lyon, Institut de Physique des 2 Infinis de Lyon,
IN2P3-CNRS--UCBL,\\
4 rue Enrico Fermi, 69622  Villeurbanne, France}
\author{Alfredo~Valcarce}
\email{valcarce@usal.es}
\affiliation{Departamento de F{\'\i}sica Fundamental and IUFFyM,
Universidad de Salamanca, 37008 Salamanca, Spain}
\date{\emph{Version of }\today}
\author{Javier~Vijande}
\email{javier.vijande@uv.es}
\affiliation{Unidad Mixta de Investigaci\'on en Radiof\'\i sica e Instrumentaci\'on 
Nuclear en Medicina (IRIMED), Instituto de Investigaci\'on Sanitaria La Fe (IIS-La Fe)-Universitat 
de Valencia (UV) and IFIC (UV-CSIC), Valencia, Spain}
\begin{abstract}
\noindent
We explore the possibility of very heavy dibaryons with three charm quarks and 
three beauty quarks, $bbbccc$, using a constituent model which should drive to 
the correct solution in the limit of hadrons made of heavy quarks. The six-body 
problem is treated rigorously, in particular taking into account the orbital, color and
spin mixed-symmetry components of the wave function. Unlike a recent claim based 
on lattice QCD, no bound state is found below the lowest dissociation threshold. 
\end{abstract}
\maketitle

\section{Introduction}
\label{se:intro}
Apart from the atomic nuclei, there is no evidence, so far, for stable multiquark 
states in the hadron spectrum. At best, there are very interesting resonances  
which lie above their lowest dissociation threshold. Sometimes the corresponding 
fall-apart decay is suppressed, and a rather narrow resonance is observed. There 
are several recent reviews, see, for instance~\cite{Briceno:2015rlt,
Chen:2016qju,Esposito:2016noz,Richard:2016eis,Lebed:2016hpi,Ali:2017jda,Ali:2019roi,Brambilla:2019esw}.
 
Several stable multiquarks have been predicted along the years, based on various 
mechanisms. The famous $H(uuddss)$ by Jaffe~\cite{Jaffe:1976yi} tentatively gets 
its binding from coherences in the chromomagnetic sector. The same mechanism is 
also at work in other configurations, such as the anticharmed 
pentaquark~\cite{Lipkin:1987sk,Gignoux:1987cn,Leandri:1989su}. 
But, when a full quark model calculation 
is performed the stability does not survive 
the breaking of flavor SU(3)~\cite{Rosner:1985yh,Richard:2019plb}, 
nor the unavoidable dilution of the structure that 
weakens the strength of the chromomagnetic terms~\cite{Oka:1988yq,Kim:2013vym}. 

Another possibility is the combination of two heavy quarks and two light antiquarks, 
$QQ\bar q\bar q$. Here, there is favorable chromomagnetic effect if $\bar q\bar q=\bar u\bar d$, 
but the novelty is the chromoelectric binding that exploits the breaking of charge conjugation 
when the quark-to-antiquark mass ratio $M/m$ departs from unity. This is the same mechanism 
that makes the hydrogen molecule much more stable than the positronium 
molecule~\cite{Richard:2016eis}. 

During the last years, lattice QCD based studies have managed to reach 
calculations at almost the physical pion mass. Thus, the HAL QCD Collaboration
has studied dibaryons containing light or strange quarks concluding the 
existence of barely bound $\Omega \Omega$~\cite{Gongyo:2017fjb}
and $N\Omega$~\cite{Iritani:2018sra} states close to the unitary limit.
Lattice calculations with heavy quarks are advantageous over the light 
counterparts because the two point correlators are less noisy and the signal to noise
is far better. 

Recently, Junnarkar and Mathur~\cite{Junnarkar:2019equ} reported 
the first lattice QCD study of dibaryons with heavy quark flavors. They evaluated deuteron-like dibaryon 
structures with the quark contents $(uuc)(ucc)$, $(ssc)(scc)$, $(uub)(ubb)$, $(ssb)(sbb)$, and $(ccb)(cbb)$ and 
quantum numbers $(I)J^P=(0)1^+$. The authors conclude that $(ssc)(scc)$, $(ssb)(sbb)$, 
and $(ccb)(cbb)$ are clearly below the two-baryon thresholds. 
It is important to note that for these systems the lowest 
two-baryon threshold is always made of two spin $3/2$ baryons 
with quark content $(QQQ)(qqq)$. The existence of such a deep bound states in
the heavy quark sector should be captured in any model having the right QCD
properties in the heavy quark sector. In particular a constituent model
approach should reflect such deep binding, if it exists~\cite{Karliner:2017qjm,Eichten:2017ffp}.

Thus, the aim of the present note is to revisit the aforementioned configurations 
in a full-fledged calculation considering the internal mixed symmetry components of 
the hexaquark wave function,
and study whether or not three $c$ quarks and three $b$ quarks do form a bound state. 
\section{Model and method}
We have calculated the energy of $bbbccc$ and its threshold using a standard constituent model. 
The main assumption, on which we shall come back in the last section, consists of adopting a 
pairwise interaction with color-octet exchange structure. 
The interaction has two terms only, i.e., spin-orbit and tensor forces are neglected:
\begin{itemize}
	\item a spin-independent or chromoelectric interaction that reads,
	\begin{equation}\label{eq:central}
	V_c=-\frac{3}{16}\sum_{i<j} (\tilde\lambda_i.\tilde\lambda_j)\,\left(-a/r_{ij}+ b\,r_{ij}\right) ~ ,\end{equation}
	\item a spin-spin or chromomagnetic interaction given by,
	\begin{equation}\label{eq:spin-spin}
	V_s=-\frac{3}{16}\,\sum_{i<j} 
	(\tilde\lambda_i.\tilde\lambda_j)\,(\vec{\sigma}_i.\vec{\sigma}_j)\frac{a_{ss}}{m_i\,m_j}\left(\frac{\mu}{\pi}\right)^{3/2}\, \exp(-\mu\,r_{ij}^2) ~,
	\end{equation}
\end{itemize}
with $a=0.2/\hbar c$, $b=0.4\,\hbar c $, $a_{ss}=2.0$, $\mu = 1.0/(\hbar c)^2$, $m_c=1.3$\,GeV, and $m_b=4.66$\,GeV.

In such potential, the ground state of three quarks corresponds to a stable baryon. A constant 
term can be introduced in \eqref{eq:central}, $-(3/16)\,C\,(\tilde\lambda_i.\tilde\lambda_j)$, but 
it would shift each baryon of the threshold by $C$ and every dibaryon by $2\,C$ and thus 
cannot modify the conclusions about the stability or instability of the latter. 

The above potential corresponds to a fit by Semay and Silvestre-Brac~\cite{Semay:1994ht}. 
It is worth to emphasize that the parameters are constrained in a simultaneous fit of
36 well-established meson states and 53 baryons, with a remarkable agreement with data, 
as can be seen in Table 2 of Ref.~\cite{Semay:1994ht}.
Notwithstanding, it has been checked that the conclusions dealing with stability or 
instability of multiquarks survive sizable variations of the parameters, and changes 
in the functional form adopted for the potential. 

The color, spin and orbital structure of the wave function has been firstly worked out in a 
basis $(123)\mhyphen (456)=(bbb)\mhyphen (ccc)$, where the Pauli principle is imposed in a more
apparent manner. The color of each cluster is either the antisymmetric 
singlet $1$ or the pair of mixed-symmetry octet $\{8_\lambda,8_\rho\}$, which are 
symmetric or antisymmetric under $1\leftrightarrow2$ (or  $4\leftrightarrow5$), respectively. 
The color decuplet states do not contribute. As for the spin, we have either 
$S_{123}$ (or $S_{456}$) = 3/2 or 1/2, which are symmetric or mixed-symmetry, 
respectively. Subsequently, the clustering $(bbc) \mhyphen (ccb)$ has also been 
used for checking purposes. In this case, the wave function has been obtained
by the transformation of the original properly antisymmetrized wave
function in the $(123)\mhyphen (456)=(bbb)\mhyphen (ccc)$ basis. The
spin-color algebra has also to be transformed into the new coupling.
Let us note that, in contrast to a deuteron-like dibaryon, the color-spin-radial wave function
must be antisymmetric due to the non-existence of a flavor-antisymmetric component.

In the ground state of the six-quark system, the orbital wave function is 
dominated by the components that are symmetric in both $bbb$ and $ccc$ sets of permutations, 
but mixed-symmetric orbital wave functions are also included. The recoupling of the spins 
is obvious. The Clebsch-Gordan of the color recoupling to an overall singlet are taken 
from \cite{Alex:2010wi}. The coupling of two or three mixed-symmetry components 
to an overall symmetric or antisymmetric state is explained, e.g., in~\cite{Richard:1992uk}. The variational 
wave functions are based on Gaussians of the type
\begin{equation}
\exp(-\varphi/2)~, \qquad \varphi=\sum_{1\le i<j\le 6} a_{ij}\,r_{ij}^2~,
\end{equation}
and appropriate combinations of their permutations~\cite{Richard:1992uk}. In practice, 
the interparticle distances $\vec r_{ij}$ are expressed in terms of standard Jacobi 
coordinates which diagonalize the intrinsic kinetic energy.
Such properly symmetrized combination of permutations also leads to 
non-diagonal Jacobi coordinate products, i.e. $\vec x_i\cdot\vec x_j$, that 
generate non-zero internal orbital angular momenta in the wave function.

It is safer to proceed by steps. For each state of total spin, from $S=3$ to $S=0$, 
we first considered symmetric orbital functions associated to color-singlet clusters, 
afterwards the admixture of color octet components coupled to mixed-symmetry 
spin states was included. In a third stage, the corrections due to mixed-symmetry orbital components 
were analyzed. In principle, especially for $S=1$ or $S=0$, the number of coupled components 
can become large, and lead to rather delicate numerical calculations. In such a case, it 
is wise to introduce the corrections one by one, and add-up the corresponding energy shifts. 
A similar strategy is used, e.g., when treating the high partial waves of the hyperspherical 
expansion~\cite{Richard:1992uk}.

\section{Results}
We first calculate the energy of the various baryons $b^n c^{3-n}$, and found without surprise 
that the lowest threshold is made of $(bbb) + (ccc)$. The convexity of the baryon spectrum 
when the masses are varied is reviewed, e.g., 
in~\cite{Richard:1992uk,Martin:1986da, Richard:1983mu,Nussinov:1999sx}. 
With the parameters of the model of Ref.~\cite{Semay:1994ht} the lowest threshold 
$\Omega_{bbb}(3/2^+) + \Omega_{ccc}(3/2^+)$ has a mass 
of 19.082~GeV. The details are given in Table~\ref{tab:baryons}.
\begin{table}[!htb]
	\caption{\label{tab:baryons}Masses, in GeV, of baryons within the constituent 
	model of Ref.~\cite{Semay:1994ht}.}
\begin{ruledtabular}
		\begin{tabular}{c|cc}
			Baryon  & $S = 3/2$ & $S=1/2$ \\
			\hline
			$bbb$   & 14.253    & $-$    \\
			$ccc$   &  4.829    & $-$    \\
			$bbc$   & 11.162    & 11.137 \\
			$ccb$   &  8.023    & 7.972  \\
		\end{tabular} 
	\end{ruledtabular}
\end{table}	

For the hexaquarks $bbbccc$, the energy was always found above the threshold, even 
when color octet and/or mixed-symmetry orbital and spin wave functions are introduced for both 
three-quark clusters. More precisely:
\begin{itemize}
\item Spin 3: As the spin wave function is fully symmetric, we have targeted an additional antisymmetric
wave function made of mixed-symmetry orbital and color components.

\item Spin 2: Besides the wave function mentioned above we included antisymmetric wave functions
made of mixed-symmetry orbital, color and spin components for one of the subclusters of three
identical quarks. Such component resembles two baryons with $S=1/2$ and $S=3/2$.

\item Spin 0 and 1: Besides the components mentioned above we included antisymmetric wave functions
made of mixed-symmetry orbital, color and spin components for both subclusters of three identical
quarks. Such component would take account of internal states made of two $S=1/2$ baryons.
\end{itemize}

The negligible contribution coming from mixed symmetry components of the wave function 
can be easily understood by calculating their energy if considered separately. 
Thus, the $S=3$ state containing mixed symmetry color and orbital
components has an energy of 19.869~GeV, to be compared to the 19.098~GeV of the color-orbital
antisymmetric-symmetric component. Similarly, the $S=1$ state containing
mixed symmetry orbital-color-spin components has an energy of 19.838~GeV, to be compared to the 
19.098~GeV of the lowest component. They are far enough to have a rather weak coupling
with the dominant state. Moreover, the adjustment of the parameters always led to very 
small and even vanishing values for the $a_{ij}$ coefficients with $i\in\{1,2,3\}$ 
and $j\in\{4,5,6\}$. This is clear signature of a converged variational calculation 
in absence of bound states.

Note that in the case of equal masses $m_b=m_c$, there is no bound state either, and this 
can be understood by an argument of symmetry breaking that was already developed for 
comparing $\bar3 3$ and $6\bar6$ tetraquarks with equal masses~\cite{Richard:2018yrm}. Let, indeed, 
\begin{equation}\label{eq:H-N}
H=\sum_{i=1}^N\frac{\vec p_i^{\, 2}}{2\,m}+\sum_{1\le i<j\le N} g_{ij}\,V(r_{ij})~,
\end{equation}
be an Hamiltonian with attractive pair potential $V$, and the cumulated strength fixed, say 
$\sum g_{ij}=N/(N-1)/2$. The energy is maximal for equal strengths $g_{ij}=1$ and is a concave 
function of these variables $g_{ij}$, as they enter $H$ linearly. So, schematically, the more 
asymmetric the $\{g_{ij}\}$ distribution, the lower the energy. Clearly, for $N=6$, one can 
hardly find a distribution more asymmetric than that of the threshold with 
$g_{12}=g_{23}=g_{13}=g_{45}=g_{56}=g_{46}\neq 0$ and the other $g_{ij}=0$,  and in the regime 
of heavy quarks such as $b$ and $c$, the hyperfine corrections play a minor role and cannot 
generate binding by themselves.  Our numerical calculations also show that the results based 
on a frozen set of color coefficients are not  significantly modified by the coupling of the 
various allowed channels. 
\section{Discussion and outlook}
The main conclusion of our accurate six-body calculation is the absence of binding for $bbbccc$ 
and similar configurations. This outcome is based on an explicit potential model and several 
variants of this potential. The principal reason for this instability is the constraint of 
antisymmetrization in both the 
$b$ and the $c$ sectors, which  prohibits the mixing of all possible configurations corresponding 
to a color-neutral hexaquark of given spin.  As a toy model, we calculated a configuration 
$bb'b'' c c' c''$ with the same masses as before and the same 
potential~\eqref{eq:central}-\eqref{eq:spin-spin},  but non-identical $b$-type and $c$-type 
of quarks. Then a binding of about 100~MeV is obtained.

A related investigation consists of questioning the prescription 
\begin{equation}\label{eq:color-add}
V=-\frac3{16}\sum_{i<j} \tilde\lambda_i.\tilde\lambda_j\,v(r_{ij})~,
\end{equation}
where the normalization is such that $v(r)$ is the quarkonium potential.
For the linear part of the interaction, say $v(r)=\sigma\,r$, a challenging alternative 
consists of building the minimal string or set of strings linking the quarks. For a baryon, 
this leads to the well-known Fermat-Torricelli construction
\begin{equation}
\label{eq:Fermat}
V=\sigma\,\min_J\left(r_{1J}+r_{2J}+r_{3J}\right)~,
\end{equation}
which gives an energy slightly higher than the additive rule~\eqref{eq:color-add}. However, 
for tetraquarks, if one generalizes~\eqref{eq:Fermat} by a combination of flip-flop and 
connected strings, namely,
\begin{gather}
\label{eq:flip-flop}
V=\sigma\min_J\left(r_{3}-r_{1}+r_{4}-r_2,
r_{3}-r_{2}+r_{4}-r_1, V_{YY}\right)~, \nonumber \\
V_{YY}=
\min_{JK}\left(r_{1J}+r_{2J}+r_{JK}+r_{3K}+r_{4K}\right)~,
\end{gather}
then the potential becomes much more favorable as compared to the 
additive rule, and binding of $QQ'\bar q\bar q'$ is obtained already 
for a mass ratio $M/m=1$. However, an important restriction has to be enforced: 
the quarks and the antiquarks should be different, i.e., not submitted to the 
Pauli principle, even so equal masses are adopted, for the sake 
of simplicity~\cite{Vijande:2013qr}. 

If the same strategy of a string confinement is adopted for six quarks, with 
the potential resulting form a minimization over various connected strings and 
flip-flop configurations, then binding is also obtained \cite{Vijande:2011im}. 
For a purely linear interaction $v(r)=r$, and a mass ratio $M/m=1$ associated 
with a light mass $m=1$, it was found that the hexaquark $QQQqqq$ is bound by 
about $0.491$ in such dimensionless units. If one restores the  appropriate scales, 
namely $m_c\simeq1.5$~GeV and a string tension $\sigma\simeq0.2\,\mathrm{GeV}^2$, 
this corresponds to about 0.1~GeV. But, again, the possibility of optimizing the 
strings continuously as the quarks move requires a full waiver of 
antisymmetrization \cite{Vijande:2011im}. In other words, a constituent model leaned 
on the string confinement predicts $bb'b'' c c' c''$ to be stable with a binding energy 
of about 100~MeV. This state disappears when the $b$ and the $c$ become identical.

To conclude, potential models, when treated seriously, do not lead to a proliferation 
of stable multiquarks. In particular, there is no evidence for any stable super-heavy 
hexaquark of the type $bbbccc$. 

\acknowledgments
This work has been partially funded by Ministerio de Econom\'\i a, Industria y Competitividad
and EU FEDER under Contracts No.\ FPA2016-77177 and RED2018-102572-T.
%
% \bibliographystyle{unsrt}
% \bibliography{hexaquark}
% \end{document}

\end{document}